\documentclass{ws-procs975x65}
\begin{document}
\title{Accelerating   Cosmology Driven by Gravitational Production of Dark Matter Particles: Constraints from SNe Ia and H(z) data}
\author{R. C. SANTOS$^{1}$, F. E. SILVA$^{2}$, J. A. S. LIMA$^{3}$}
\address{$^{1}$Departamento de Ci\^encias Exatas e da Terra, Universidade Federal
de S\~ao Paulo, 09972-270  Diadema, SP, Brasil, e-mail: cliviars@gmail.com \\$^{2}$Escola de Ci\^encias \& Tecnologia, Universidade Federal do Rio Grande do Norte, \\
59078-970, Natal, RN, Brazil, e-mail: edsonsilva@ect.ufrn.br \\$^{3}$Departamento de Astronomia, Universidade de S\~{a}o
Paulo \\ Rua do Mat\~ao, 1226 - 05508-900, S\~ao Paulo, SP, Brazil, e-mail:limajas@astro.iag.usp.br }
\begin{abstract}
The free parameters of a flat accelerating model without dark energy are constrained by using Supernovae type Ia  and observational H(z) data.
Instead of the vacuum dominance, the present accelerating stage in this modified Einstein-de
Sitter cosmology is a consequence of the gravitationally-induced particle
production of cold dark matter.  The model present a transition from a decelerating to an accelerating regime at
low redshifts,  and is  also  able to harmonize a cold dark matter picture with
the latest measurements of the Hubble constant $H_0$,  the Supernovae observations (Constitution  sample), and  the H(z) data.
\end{abstract}
 \bodymatter

\section{Introduction}
A growing body of complementary cosmological data are suggesting that the
Universe underwent a late time  transition from a decelerating to an accelerating
expansion~\cite{accel,Hicken,WMAP7}. Current data are accurately fitted by a flat
FRW type cosmology containing nonrelativistic matter  plus some sort of dark energy \cite{WMAP7,darkenergy}.
The simplest and by far the most popular dark energy candidate is represented by a cosmological
constant. In the so-called $\Lambda$CDM model, the cosmic
fluid contains radiation, baryons, cold dark matter plus a vacuum
energy. Nevertheless, this $\Lambda$CDM model is plagued with some difficulties like the
cosmological constant and coincidence problems.

On the other hand, the presence of a negative pressure is the key
ingredient required to accelerate the expansion. This kind of stress
occurs naturally in many different contexts when the physical
systems depart from thermodynamic equilibrium states\cite{LL}.
In this connection, as pointed out by some authors\cite{Pri}, the process of  cosmological particle
creation at the expense of the gravitational field can
phenomenologically be described by a  negative pressure, and, more interestingly, can accelerate the Universe\cite{lss,SSL,lima2010}.

In this context, we constrain the free parameters of a flat accelerating CDM
cosmology recently proposed in the literature\cite{lss,SSL}.  As we shall see, this extended CDM model is
consistent with the SNe Ia (Constitution sample)\cite{Hicken} and H(z) data \cite{DadosHz}. In addition, there is a transition from a decelerating to an accelerating regime at  redshift of the order of a few, and  the Hubble constant does not need to be small in order
to solve the age problem. Such a transition happens even if the
matter creation is negligible during the radiation and considerable
part of the matter dominated phase. In certain sense, the
coincidence problem of $\Lambda$CDM model is replaced here by a
gravitational particle creation process at low redshifts.

\section{Decelerating Parameter,  Supernova and H(z) Bounds}

For simplicity, let us consider that the spacetime is filled only by a cold dark matter component. In this case,  the Hubble and the decelerating parameters are given by: \cite{lss}

\begin{equation}
H(z)=H_0\left[\frac{\gamma
+(1-\gamma-\beta)(1+z)^{\frac{3}{2}(1-\beta)}}{1-\beta}\right],\label{Hz}
\end{equation}

\begin{equation}
q(z)=\frac{1}{2}
\left[\frac{(1-3\beta)(1-\gamma-\beta)(1+z)^{\frac{3}{2}(1-\beta)}-2\gamma}{(1-\gamma-\beta)(1+z)^{\frac{3}{2}(1-\beta)}+\gamma}\right].
\end{equation}

Note that if $\gamma =0$ there is no transition from a decelerating to an accelerating regime.  The Universe is always decelerating or accelerating depending on the value of the $\beta$ parameter.  The existence of a transition redshift depends exclusively on the $\gamma$ parameter. However, this fact does not remain  true when baryons are included \cite{SSL}.

In Figure 1a, we display the effect of the  free parameters ($\gamma, \beta$) in the reduced Hubble-Sandage diagram for the Constitution sample \cite{Hicken}.  The lowest  (yellow) curve is the prediction of the Einstein-de Sitter model.

In Figure 1b, we show the contours of constant likelihood (68.3\%,
95.4\%, and 99.7\% C.L.) in the ($\gamma, \beta$) plane  from a
$\chi^{2}$ statistics based on the Constitution set. Following
standard lines, we have marginalized our likelihood function over
the nuissance parameter $h$ ($H_0 = 100h Km.s^{-1}.Mpc^{-1}$). It is
found that the free parameters fall on the intervals $0.22 \leq
\gamma \leq 0.67$ and $0 \leq \beta \leq 0.38$ at $68.3\%$ of
confidence level. The best fit occurs for values of $\gamma=0.64$
and $\beta=0$ with $\chi^{2}_{min}=466.97$ and $\nu=395$ degrees of
freedom. The reduced $\chi^{2}_{r}=1.18$ where
($\chi^{2}_{r}=\chi^{2}_{min} /\nu$), thereby showing that the model
provides a very good fit to these data.

\begin{figure*}[h!]
\centerline{\psfig{figure=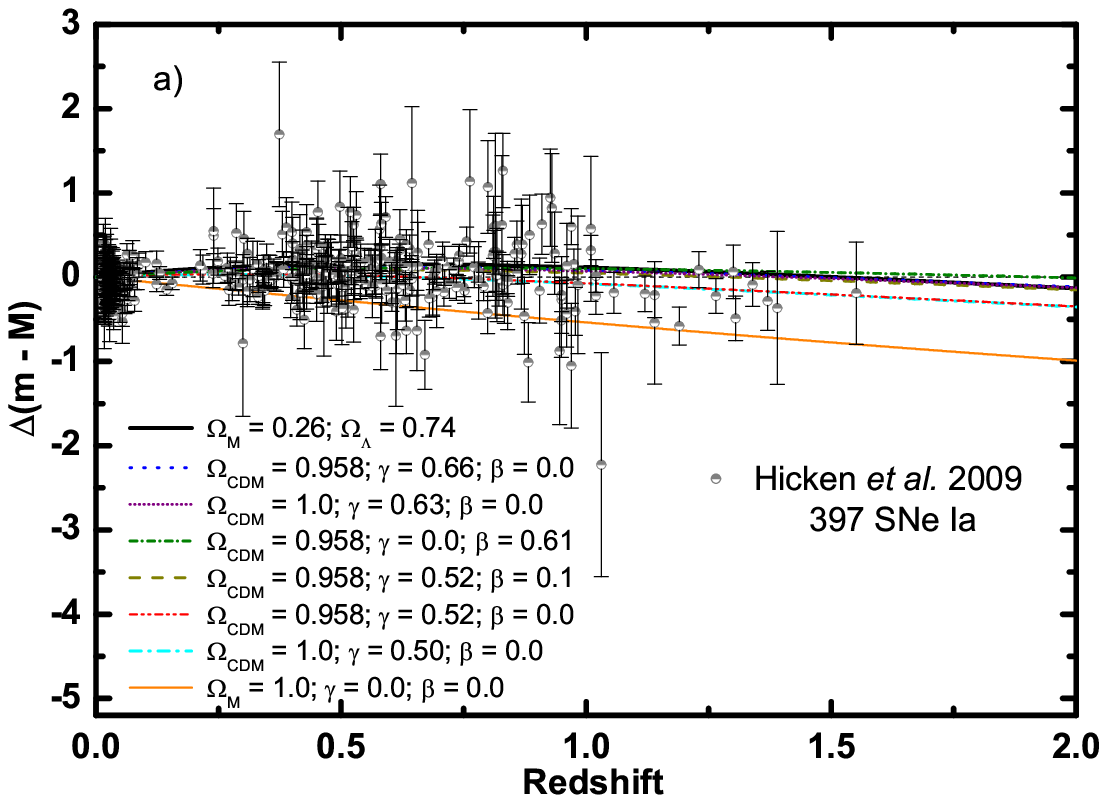,width=1.75truein,height=1.8truein}
\psfig{figure=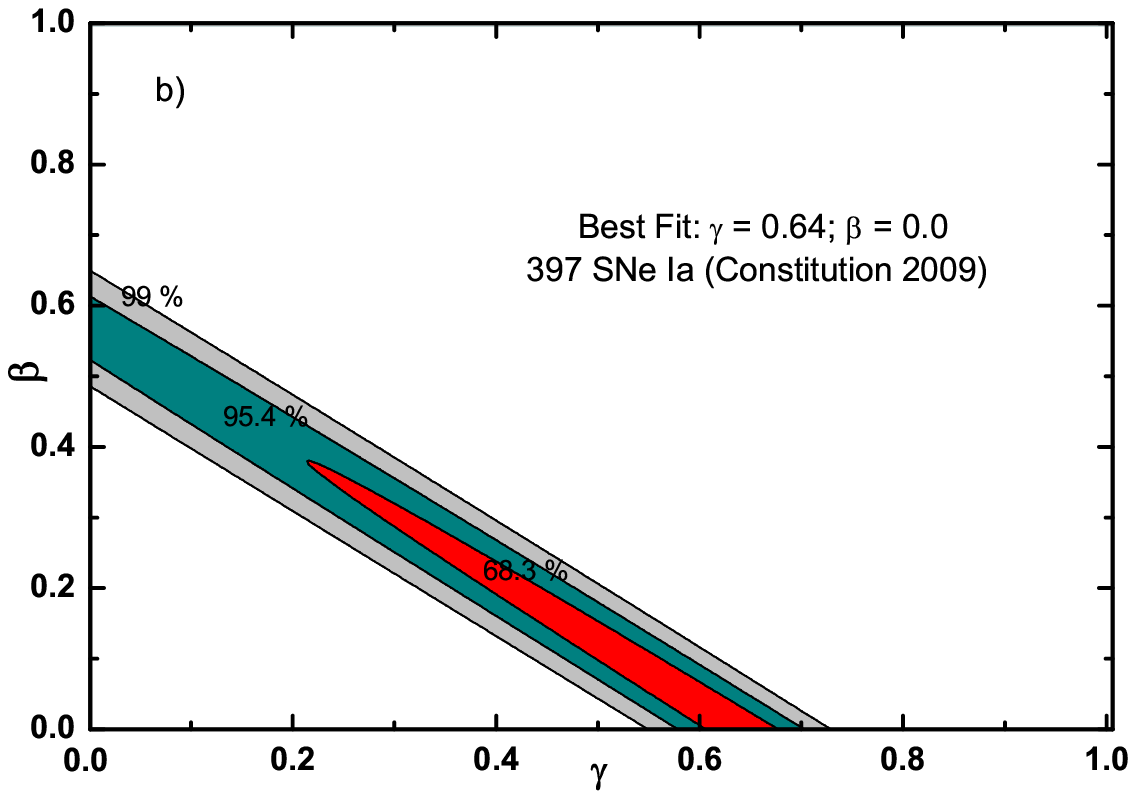,width=1.8truein,height=1.8truein}
\psfig{figure=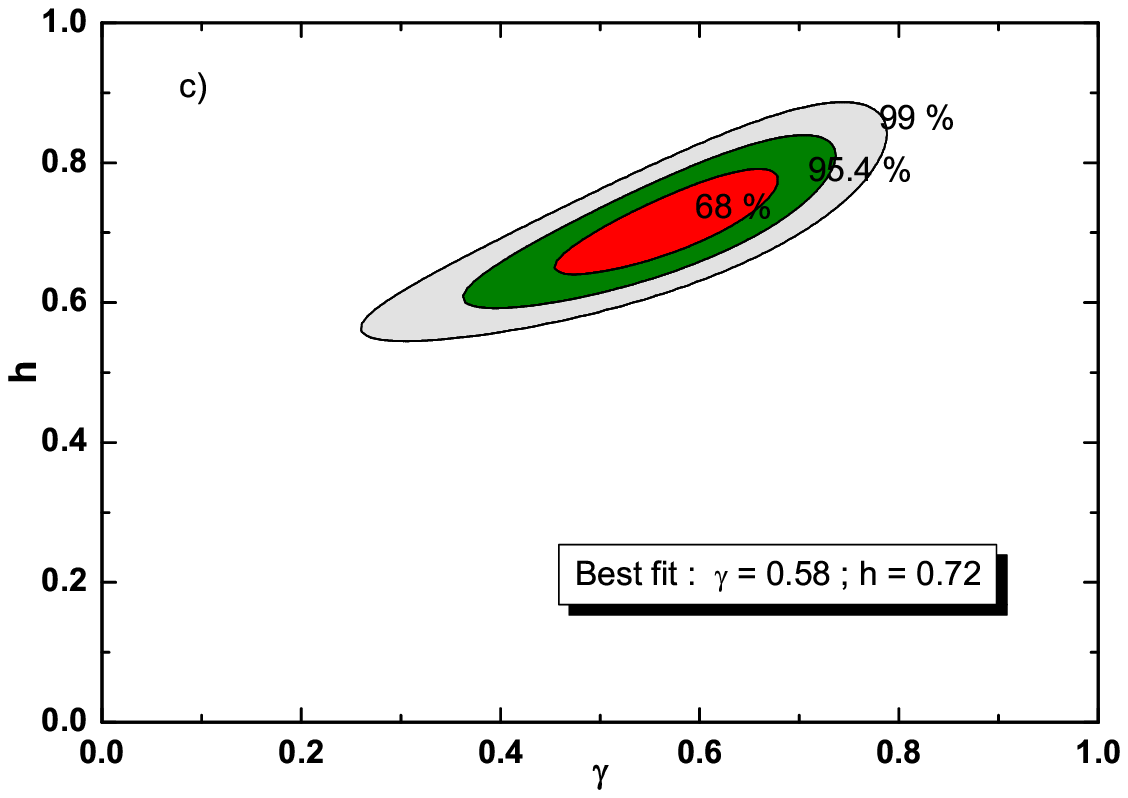,width=1.8truein,height=1.8truein}
 \hskip0.1in} \vspace{0.5cm} \caption{ {\bf{a)}}The relative
distance modulus versus redshift relations for a
variety of models with and without particle creation. {\bf{b)}}The $\gamma$-$\beta$ plane
for a flat CDM model with gravitational particle creation obtained
from the same sample.  {\bf{c)}} Confidence regions on the $\gamma$ - $h$
plane. The constraints are  $0.61 \leq h \leq 0.86$ and $0.38 \leq \gamma \leq
0.76$ ($2\sigma$). The corresponding best fits are indicated in the figures.}
\end{figure*}

In Figure 1c, we show the contours on the ($\gamma$, h) plane using the H(z) data from Stern {\it{et al.}}\cite{DadosHz}.
The free parameters are constrained by  $0.61 \leq h \leq 0.86$ and $0.38 \leq \gamma \leq
0.76$ (at $2\sigma$ C. L.). Note that the constraints on the $\gamma$ parameter are consistent with each other for these two different classes of data.

\section{Conclusion}

By using SNe Ia and $H(z)$ data,  we have discussed some constraints
on a flat accelerating cold dark matter cosmology without dark
energy.  The accelerating regime is powered by a negative pressure
associated to the gravitationally-induced creation of CDM particles.
The transition from a decelerating to an accelerating regime at late
times happens even if the matter creation is negligible during the
radiation and considerable part of the matter dominated phase (this
is equivalent to take $\beta = 0$ in all the expressions). In this
case, like in the flat $\Lambda$CDM, there is just one free
parameter, and the resulting model provides an excellent fit to the
observed dimming of distant SNe Ia data. More important, such
constraints are compatible with the latest determinations of the
Hubble constant $H_0$ (see Figure 1c). Naturally, complementary
tests must still be investigated to see whether the present scenario
may provide a realistic description of the observed Universe.

\end{document}